\documentclass[twocolumn,preprintnumbers,amsmath,amssymb,superscriptaddress]{revtex4}

\usepackage{graphicx}

\usepackage{amsmath}
\usepackage{amssymb}
\usepackage{dcolumn}
\usepackage{bm}
\usepackage{upgreek}

\begin{document}
\preprint{}
\title{Short-Pulse Amplification by Strongly-Coupled Stimulated Brillouin Scattering}

\author{Matthew R. Edwards}
\email{mredward@princeton.edu}
\affiliation{Department of Mechanical and Aerospace Engineering, Princeton University, Princeton, New Jersey 08544, USA}
\author{Qing Jia}
\affiliation{Department of Astrophysical Sciences, Princeton University, Princeton, New Jersey 08544, USA}
\author{Julia M. Mikhailova}
\affiliation{Department of Mechanical and Aerospace Engineering, Princeton University, Princeton, New Jersey 08544, USA}
\author{Nathaniel J. Fisch}
\affiliation{Department of Astrophysical Sciences, Princeton University, Princeton, New Jersey 08544, USA}

\date{\today}

\begin{abstract}
We examine the feasibility of strongly-coupled stimulated Brillouin scattering as a mechanism for the plasma-based amplification of sub-picosecond pulses. In particular, we use fluid theory and particle-in-cell simulations to compare the relative advantages of Raman and Brillouin amplification over a broad range of achievable parameters. 
\end{abstract}

\maketitle
\section{Introduction}
Parametric amplification of light in plasmas \cite{Malkin1999} offers a route to short laser pulses with ultra-high intensities far beyond the damage thresholds of solid-state devices \cite{Strickland1985}. Studies of plasma amplification have primarily considered stimulated Raman \cite{Malkin1999} and Brillouin \cite{Andreev2006} backscattering, which are based on three-wave coupling of pump and probe laser beams with Langmuir or ion-acoustic plasma waves, respectively. Amplification by stimulated Raman scattering (SRS) offers higher growth rates and shorter compressed pulses, and a large volume of work has been devoted to understanding its limiting factors, including wavebreaking \cite{Malkin1999,Malkin2014EPJST,Toroker2014,Edwards2015}, Landau damping \cite{Hur2005,Malkin2007,Yampolsky2009,Malkin2010,Yampolsky2011,Strozzi2012,Wu2014,Depierreux2014}, spontaneous Raman scattering \cite{Malkin1999,Malkin2000,Malkin2014EPJST,Malkin2000detuned,malkin2000stimulated,Solodov2004}, plasma inhomegeneities \cite{Solodov2003}, and relativistic non-linearities \cite{Malkin1999,Malkin2014EPJST,Fraiman2002,Malkin2007,Malkin2012,Lehmann2014,Malkin2014pop}. A series of experiments have demonstrated the viability of the mechanism \cite{Ping2000,Ping2002,Ping2004,Balakin2004,Cheng2005,Kirkwood2007,Ren2008,Pai2008,Yampolsky2008,Ping2009,Turnbull2012,Vieux2011,Depierreux2014}, though maximum achieved intensities are somewhat below the theoretical limits.

Stimulated Brillouin scattering (SBS) in plasma, which plays an important role in inertial confinement fusion \cite{Labaune1997interplay,Neumayer2008}, was first considered for plasma-based pulse amplification in the weakly-coupled regime \cite{Milroy1977,Milroy1979,Andreev1989,Capjack1982plasma}, where the dynamics of the ion-acoustic wave govern the interaction. SBS differs from SRS in that the frequency of ion-dependent waves is much lower than that of Langmuir waves, so the matching conditions for frequency ($\omega_\textrm{pump} = \omega_\textrm{seed} + \omega_\textrm{plasma}$) and wavevector ($\mathbf{k}_\textrm{pump} = \mathbf{k}_\textrm{seed} + \mathbf{k}_\textrm{plasma}$), or conservation of energy and momentum, dictate that the the frequency shift between the pump and seed is small. Interest in SBS has been driven by the greater robustness that this property may provide with respect to plasma inhomogeneities and because it allows the pump and seed to be produced by sources of the same wavelength  \cite{Milroy1977,Milroy1979,Capjack1982plasma,Andreev1989,Andreev2006,Lancia2009experimental,Lancia2010,Lehmann2012,Weber2013,Riconda2013,Alves2013,Lehmann2013,Riconda2013kinetic,Humphrey2013,Trines2014,Guillaume2014demonstration,Riconda2014plasma,Frank2014amplification,Schluck2015,Lehmann2015,Shoucri2015,Lehmann2016temperature,Lancia2016signatures,Chiaramello2016optimization} or even the same beam \cite{Peng2016Single}. The disadvantage of weakly-coupled SBS is that the minimum duration of the amplified pulse is set by the period of the ion-acoustic wave, which is long compared to the pulse durations that can be achieved with Raman amplification. The strongly-coupled stimulated Brillouin scattering (SC-SBS) regime \cite{Andreev2006}, where, at larger pump intensities and plasma densities, the ion-acoustic wave becomes a higher-frequency driven quasi-mode, allows shorter pulse durations and may be suitable for amplification of picosecond and sub-picosecond pulses. Previous work has suggested that SC-SBS allows compression to durations within an order of magnitude of those for Raman amplified pulses \cite{Andreev2006}. 

Initial study of amplification by SC-SBS considered plasma densities above one quarter of the critical density, where SRS is suppressed, but in an effort to avoid deleterious instabilities, particularly filamentation, study has expanded to include lower plasma densities, where SBS is in direct competition with SRS \cite{Lancia2010,Weber2013,Riconda2013,Trines2014}. In this regime, the dominant amplification mechanism may not be obvious \cite{Jia2016}, and some care must be taken to determine the underlying process. Other works have considered relativistic \cite{Lehmann2012}, thermal \cite{Lehmann2016temperature}, and collisional \cite{Humphrey2013} effects, as well as the role of instabilities like forward Raman scattering (FRS) \cite{Trines2014} on the amplification process. SBS has been demonstrated experimentally \cite{Lancia2010,Guillaume2014demonstration,Lancia2016signatures}, with recently published work suggesting amplification in the self-similar regime, where depletion of the pump beam is significant \cite{Lancia2016signatures}.  

In principle either SRS or SBS may be useful for amplification of picosecond or sub-picosecond pulses to near-relativistic intensities, but the question of which process is more appropriate under specific conditions remains to be resolved. In this paper, we draw on analytic models and particle-in-cell (PIC) simulations to examine the relative advantages of Brillouin amplification over Raman amplification for the generation of ultra-high-intensity short laser pulses, with particular emphasis on clarifying the usefulness of SBS for amplifying sub-picosecond pulses. 

The paper is structured as follows. In Sec.~\ref{sec:theory}, we discuss the mechanism of Brillouin amplification, taking particular care with the definitions of weak-coupling and strong-coupling. Section \ref{sec:RvB} compares the relative advantages of Raman and Brillouin amplification over a broad range of parameters. In Sec.~\ref{sec:lims}, we consider the limits of SBS, and Sec.~\ref{sec:collisional} covers a collisional regime where SBS may be considerably more attractive than SRS. The conclusions of this work are summarized in Sec.~\ref{sec:conc}.

\section{Theory of Brillouin Amplification}
\label{sec:theory}

To understand the mechanism of Brillouin amplification and provide a solid basis for discussion of its merits, we will start with a simple physical description of the process, followed by a detailed analytic derivation of the dispersion relation in the weakly- and strongly-coupled regimes.

The physics of stimulated Brillouin scattering depend strongly on the intensity of the pump (represented by the normalized pump strength $a_0 = E_0/E_\textrm{rel}$, where $E_\textrm{rel}= m_e \omega_L c / e$, $E_0$ is the maximum electric field, $\omega_L$ is the laser frequency, and $e$ and $m_e$ are the charge and mass of an electron), and the plasma density ($N = n_e/n_c$ where $n_c = m_e \omega_L^2 / 4 \pi e^2$ is the plasma critical density and $n_e$ is the electron number density). SBS is also affected by the electron temperature ($T_e$), the mass ratio $Z m_e / m_i$, where $Z$ is the ion charge and $m_i$ is the ion mass, and, to a lesser extent, the ion temperature ($T_i$). Here, we focus on the regime where the pump is not relativistic ($a_0 \ll 1$), and we note that non-relativistic SBS is only possible for underdense plasmas ($N < 1$). We will also consider the ion temperature to be negligible throughout ($T_i \ll T_e$), noting that violation of this condition will tend to decrease the performance of SBS, and we take $Z = 1$ everywhere.

\subsection{The Physics of Brillouin Scattering}
Stimulated Brillouin scattering is a three-wave coupling interaction of two electromagnetic waves with an ion-acoustic wave. In a light field with a slowly varying envelope, e.g. that produced by counterpropagating laser beams, charged particles are driven to regions of lower intensity by the ponderomotive force:
\begin{equation}
F_p = - \frac{e^2}{4 m \omega^2} \nabla  \left( E^2 \right)
\end{equation}
where $\omega$ is the high frequency oscillation, $E$ is the electric field strength, and $e$ and $m$ are the particle charge and mass. Although the ponderomotive force acts on both ions and electrons, fluctuations in ion density are not produced by direct ponderomotive forcing of the ions; rather, the ions respond to the electrostatic field of displaced electrons. The ponderomotive force acting on the ions is smaller than that acting on the electrons by a factor of $m_i/m_e$. Since the electrostatic force on the ions is equal to the electrostatic force on the electrons, which should balance the ponderomotive force on the electrons, the electrostatic force on the ions is larger than the direct ponderomotive force by a factor of $m_i/m_e$. The result of this ponderomotive forcing is fluctuations in the ion and electron densities at a frequency matched to the difference between the pump and seed laser frequencies. 

In general, the ion-acoustic wave is influenced by both electrostatic and pressure effects, but for SBS the electrostatic term generally dominates and the pressure term may be neglected. The resulting electron density fluctuations modulate the transverse current induced by the pump, scattering energy into the seed and producing an amplified seed pulse with central frequency down-shifted by the ion wave frequency. Since the seed drives density fluctuations more strongly as it grows, both the seed and the density fluctuations increase in time. SBS is limited in duration by the period of the ion-acoustic waves, or in the case of SC-SBS the period of the quasi-mode, since the description of the density fluctuation as a wave is only valid if the wavelength is shorter than the regime under consideration. The growth of other instabilities limits the maximum intensities and minimum pulse durations that can be achieved in practice. For example, seed pulses with intensities higher than $10^{17}-10^{18}$ W/cm$^2$ will break up due to relativistic self-modulation. 

\subsection{The Ion-Acoustic Dispersion Relation}
\label{sec:disp}
To understand the properties of Raman and Brillouin amplification, we use a linear analysis of the two-fluid plasma model, a common approach in the literature \cite{Kruer2003,Guzdar1996stimulated,Forslund1975}. Though this method is not strictly valid in all regimes of interest, and in particular is not appropriate where kinetic effects are important, where the pump is significantly depleted, or where fluctuating quantities are large, it provides a simple and reasonably accurate estimate of the frequency and growth rate associated with both the Raman and Brillouin instabilities. The properties of Raman and Brillouin amplification presented in the literature are often based on formulas derived from this type of analysis, so despite the limits of this fluid model in some regimes of interest, it will be our starting point for the comparison of Brillouin and Raman amplification. 

Both Raman and Brillouin amplification are a result of light scattering from electron density fluctuations \cite{Kruer2003}:
\begin{equation}
\left(\partial^2_t - c^2 \partial^2_x + \omega_{pe}^2 \right) \tilde{\mathbf{A}}_S = - \frac{4 \pi e^2}{m} \tilde{n}_e \mathbf{A}_L
\label{eqn:light}
\end{equation}
where $\tilde{\mathbf{A}}_S$ is the vector potential of the scattered light, $\mathbf{A}_L$ is the vector potential of the pumping light and $\tilde{x}$ marks fluctuating quantities, i.e. $n_e = \tilde{n}_e + n_0$. For $m_i \gg m_e$, the ions do not contribute directly to the scattering even for SBS because the high ion mass results in a low transverse ion quiver velocity and a correspondingly negligible contribution to the transverse current.

The dynamics of the electron density fluctuations differ between Raman and Brillouin scattering. For SBS, the electron density fluctuation is associated with an ion wave, with dynamics given by:\begin{equation}
\left(\partial^2_t - c_s^2 \partial_x^2 \right) \tilde{n}_e = \frac{Z e^2 n_0}{m_e m_i c^2} \partial_x^2 \left(\mathbf{A}_L \cdot \tilde{\mathbf{A}}_S \right)
\label{eqn:dens}
\end{equation}
where $c_s = \sqrt{Z T_e/m_i}$ is the speed of sound for negligible ion temperature. For comparison, the density fluctuations in SRS are tied to Langmuir waves, given by:
\begin{equation}
\left(\partial^2_t + \omega_e^2  - 3 v_e^2\partial_x^2 \right) \tilde{n}_e = \frac{e^2 n_0}{m_e^2 c^2} \partial_x^2 \left(\mathbf{A}_L \cdot \tilde{\mathbf{A}}_S \right)
\end{equation}
where $v_e =  \sqrt{T_e/m_e}$ is the electron thermal speed. In a plasma where the two species have comparable masses, e.g. an electron-positron plasma, the dynamics of both species must simultaneously be considered \cite{Edwards2016}, but when $m_i \gg m_e$, we may treat the modes separately and neglect direct contributions from the ion current. 

Combination of Eqs.~\ref{eqn:light} and \ref{eqn:dens} leads to a general expression for the dispersion relation of Brillouin scattering \cite{Kruer2003,Guzdar1996stimulated}:
\begin{equation}
\left[\omega^2 - c_s^2 k^2\right]\left[\omega^2 - 2\omega_0 \omega -c^2 k^2 + 2c^2 k_0 k\right] = \frac{k^2 v_\textrm{osc}^2 \omega_{pi}^2}{4}
\end{equation}
where $\omega$ and $k$ are the frequency and wavenumber of the plasma wave and $\omega_0$ and $k_0$ are those of the pump laser. The electron oscillation velocity is $v_\textrm{osc} = a_0 c$ and $\omega_{pi} = \sqrt{4 \pi n_i Z e^2 / m_i}$ is the ion plasma frequency. We note that $\omega \ll \omega_0$ for all modes of interest, so the $\omega^2$ term in the second bracket may be dropped. Letting $\tilde{\omega} = \omega/kc_s$, $k = 2k_0 - \delta k$ and $\delta \tilde{k} = c \delta k / 2 k_0 c_s$, we simplify this equation to:
\begin{equation}
\left[\tilde{\omega}^2 - 1\right] \left[ \tilde {\omega} - \delta \tilde{k} \right] = - \Lambda 
\label{eqn:dispsimp}
\end{equation}
where we have taken $\delta k \ll 2k_0$ and
\begin{equation}
\Lambda = \frac{\omega_{pi}^2 v_{\textrm{osc}}^2}{16 k_0 c_s^3 \omega_0} = \frac{\omega_{pe}^2 v_{\textrm{osc}}^2}{16 k_0 c_s v_e^2 \omega_0}
\end{equation}
or equivalently:
\begin{equation}
 \Lambda = \left(\frac{a_0}{4}\right)^2 \frac{N}{\sqrt{1-N}} \left(\frac{m_i}{Z m_e}\right)^{\frac{1}{2}} \left(\frac{c}{v_e}\right)^3
\end{equation}
Note that in the last equation we have used $N = \omega_{pe}^2/\omega_0^2$ and $k_0 = \omega_0 \sqrt{1-N} / c$.

\begin{figure}
\centering
\includegraphics[width=\linewidth]{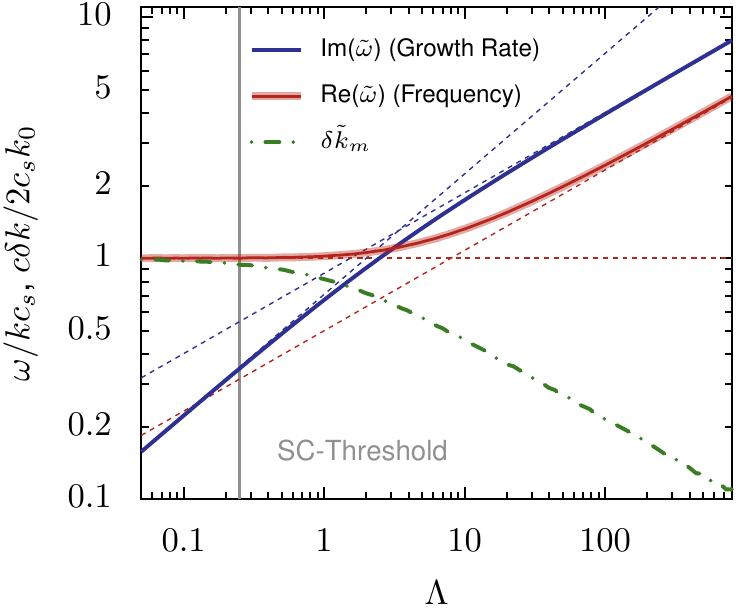}
\caption{Frequency and growth rate of Brillouin scattering in the transition between the weakly-coupled ($\Lambda \ll 1$) and strongly-coupled ($\Lambda \gg 1$)  regimes. The value of $\delta \tilde{k}$ which produces the largest growth rate is also shown. The strongly-coupled and weakly-coupled limits of both the growth rate and frequency are marked by thin dashed lines. The vertical line marks the conventional strong-coupling threshold.}
\label{fig:SCBalpha}
\end{figure}

The wavenumber adjustment which produces the maximum Brillouin growth rate, $\delta \tilde{k}_m (\Lambda)$, is $1$ in the weakly-coupled limit \cite{Kruer2003} and $0$ in the strongly-coupled limit \cite{Guzdar1996stimulated}. The value of $\delta \tilde{k}_m$ in the intermediate regime may be found numerically. With the value of $\delta \tilde{k}_m$ determined, the growth rate and frequency associated with the Brillouin mode may be calculated for arbitrary $\Lambda$ using Eq.~\ref{eqn:dispsimp}, as shown in Fig.~\ref{fig:SCBalpha}. The value of $\delta \tilde{k}_m(\Lambda)$ is also shown, and the weakly- and strongly-coupled SBS growth rate and frequency asymptotes are marked with dashed lines. 

Equation \ref{eqn:dispsimp} may be solved analytically in the strongly- and weakly-coupled limits. In the weakly-coupled limit, $\Lambda \ll 1$, so $\tilde{\omega} - 1 \ll 1$ and 
\begin{equation}
\tilde{\omega} - 1  = i \sqrt{\frac{\Lambda}{2}}
\end{equation}
or, equivalently:
\begin{equation}
\omega = c_s k + i \frac{1}{2\sqrt{2}} \frac{v_\textrm{osc} w_{pi} k_0}{\sqrt{\omega_0 c_s k_0}}
\end{equation}
For strongly-coupled Brillouin scattering, $\Lambda \gg 1$, so the solution requires $\tilde{\omega} \gg 1$. Since $\tilde{\omega} \gg \delta \tilde{k}$, justified because $\delta \tilde{k}_m < 1$, Eq.~\ref{eqn:dispsimp} immediately simplifies to:
\begin{equation}
\tilde{\omega} = \left(-\Lambda \right)^{1/3}
\end{equation}
Which can be written in the usual form as:
\begin{equation}
\label{eqn:scb}
\omega = \left(\frac{1}{2} + i\frac{\sqrt{3}}{2}\right)\left[\frac{v_\textrm{osc}^2 k_0^2 \omega_{pi}^2}{2 \omega_0}\right]^{1/3}
\end{equation}
noting that this results from choosing a particular root of $(-1)^{1/3}$. 

The strongly-coupled Brillouin condition derived by Forslund \cite{Forslund1975} which divides these two regimes:
\begin{equation}
\label{eqn:threshold}
\left(\frac{v_\textrm{osc}}{v_e}\right)^2 > \frac{4 k_0 c_s \omega_0}{\omega_{pe}^2}
\end{equation}
may be expressed simply in terms of $\Lambda$ as:
\begin{equation}
\Lambda > \frac{1}{4}
\end{equation}

It is clear from Fig.~\ref{fig:SCBalpha} and Eq.~\ref{eqn:dispsimp} that the regime of SBS is determined entirely by the value of $\Lambda$. From Fig.~\ref{fig:SCBalpha}, we also note that though the classical threshold (Eq.~\ref{eqn:threshold}) marks the edge of the weakly-coupled regime, it is necessary to satisfy the stricter condition $\Lambda \gg 1$ for the strongly-coupled equations to be valid. This more stringent condition requires $a_0$ to be about an order of magnitude larger than the requirement set by Eq.~\ref{eqn:threshold}, since the intermediate solution does not converge to the strongly-coupled asymptote until $\Lambda \approx 25$.

\begin{figure*}[tb]
\centering
\includegraphics[width=\linewidth]{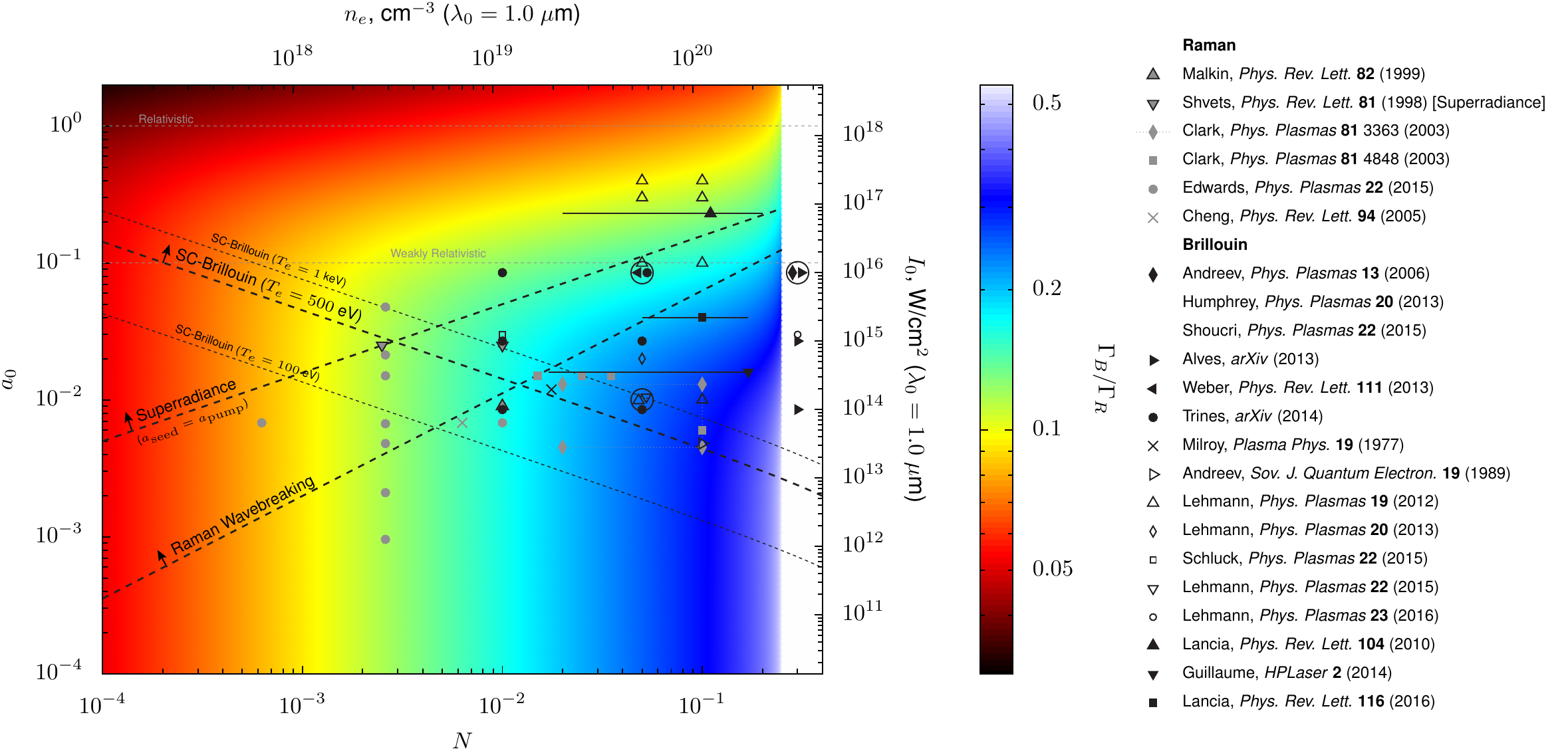}
\caption{Ratio of the stimulated Brillouin scattering growth rate ($\Gamma_B$) to the stimulated Raman scattering growth rate ($\Gamma_R$) as a function of laser field strength ($a_0$) and plasma density ($N$). Note that Raman scattering is not possible for $N \ge 0.25$, so the ratio goes to infinity. The black dashed lines labeled SC-Brillouin mark $\Lambda = 1/4$, i.e. the conventional threshold for strongly-coupled Brillouin scattering, at different temperatures. The growth rate ratios in this plot are calculated for an electron temperature $T_e = 500$ eV and an ion mass $m_i = 3600 m_e$. These are common choices in the literature, though the studies presented here include a range of electron temperatures and mass ratios. For example, recent experimental work \cite{Lancia2016signatures} was conducted with hydrogen ($m_i = 1836 m_e$) at an estimated temperature $T_e = 100$ eV. Literature points are drawn from references \cite{Edwards2015,Clark2003,Clark2003a,Shvets1998,Cheng2005,Malkin1999,Weber2013,Andreev2006,Trines2014,Lancia2010,Alves2013,Milroy1977,Andreev1989,Lehmann2013,Schluck2015,Lehmann2015,Lehmann2012,Humphrey2013,Shoucri2015,Guillaume2014demonstration,Lancia2016signatures,Lehmann2016temperature}. In certain references the plasma profile and pump amplitudes are non-uniform; the indicated points represent the maximum density and pump amplitude of the interaction. The derived growth rates are based on a linear analysis of the fluid model set of governing equations, and thus may not be strictly valid in regimes where kinetic or non-linear effects are important. Multiple markers are circled when they all represent the same conditions. The superradiant threshold depends on the product $a_{\textrm{seed}}a_{\textrm{pump}}$, so the threshold will move for different ratios of seed to pump strength. For experimental work \cite{Lancia2010,Guillaume2014demonstration,Lancia2016signatures} the full range of tested densities is marked by a line.}
\label{fig:GamBtoR}
\end{figure*}

In Fig.~\ref{fig:GamBtoR} the numerically found values of the SBS growth rate are compared to the Raman growth rate \cite{Kruer2003}:
\begin{equation}
\Gamma_R = \frac{k v_\textrm{osc}}{4} \left[\frac{\omega_{pe}^2}{\omega_{ek}(\omega_0 - \omega_{ek})}\right]^{1/2}
\end{equation}
where $\omega_{ek}^2 = \omega_{pe}^2 + 3v_e^2k^2$, as a function of laser intensity and plasma density at $T_e = 500$ eV and $m_i/m_e = 3600$. The change in scaling near the strongly-coupled threshold is readily apparent. Note that Brillouin scattering only has a larger growth rate in regions where Raman scattering is suppressed, i.e. $N \ge 0.25$, and the fundamental equations lose validity when relativistic ($a_0 > 0.1$) or kinetic (e.g. strongly wavebreaking) effects are important, so the ratios in this figure are only roughly correct. Figure \ref{fig:GamBtoR} also shows the locations of previous studies of Brillouin amplification in density-intensity space, along with examples of parameters in Raman-amplification studies. For work conducted at wavelengths other than 1 $\mu$m, only the values of $a_0$ and $N$ correspond to what was previously published, and some of the presented studies chose different values of $T_e$ and $m_i/m_e$, so that they fall in different places with respect to the strongly-coupled thresholds. For cases using plasma density gradients or pump pulses with Gaussian envelopes, the maximum values have been marked. Superradiant-type amplification additionally requires that the seed pulse be sufficiently short and intense, so it may not be observed even if the density and pump intensity requirements are fulfilled.

\section{Comparing Raman and Brillouin Amplification}
\label{sec:RvB}

The first question facing the designer of a plasma-based amplifier is whether the amplification mechanism should be Raman or Brillouin scattering. Since the advantage of plasma amplification over well-developed solid state systems is the high intensities which can be reached, the primary application for plasma amplifiers is generating ultra-high intensity, ultra-short laser pulses. An ideal system should be stable, robust, and be relatively flexible with respect to frequency, intensity, duration, and polarization. The difficulty of generating a large truly uniform plasma requires that the amplification mechanism be reasonably resilient to inhomogeneities in temperature or density. 

The differences between SRS and SBS have been articulated with varying degrees of rigor, though direct comparisons appear mostly in literature on SBS, where the following advantages for Brillouin amplification over Raman amplification have been presented: 
(1) the pump and seed lasers may have almost the same frequency \cite{Milroy1979,Andreev2006,Lehmann2013,Weber2013,Riconda2013kinetic,Shoucri2015,Lancia2016signatures}, 
(2) energy loss to the plasma wave, which results from conservation of energy and is described by the Manley-Rowe relations, may be lower for SBS than for SRS, \cite{Milroy1979,Andreev2006,Riconda2013,Guillaume2014demonstration}, i.e. a greater degree of pump depletion is obtained \cite{Weber2013,Riconda2013,Shoucri2015}, 
(3) SBS is more robust than SRS to plasma inhomogeneities in density or temperature \cite{Milroy1979,Andreev2006,Weber2013,Trines2014}, 
(4) SBS is better suited for producing pulses with high total power or energy, in part because the lower sensitivity to inhomogeneity allows larger diameter plasmas to be used \cite{Trines2014}, 
(5) only SBS may be used in the regime $0.25 < N < 1$ \cite{Guillaume2014demonstration},
(6) the duration of a Brillouin-amplified pulse can be shortened to within a factor of 8 of that for a Raman compressed pulse \cite{Andreev2006}, suggesting that  the two methods are capable of  comparable pulse-compression,
(7) a shorter interaction length is required for SBS because the energy transfer is fast \cite{Andreev2006,Weber2013,Shoucri2015}, which is sometimes quantified as SRS requiring mm to cm scale plasmas whereas SBS can be conducted in 100 $\mu$m \cite{Riconda2013}, and
(8) SBS may be viable in regimes where SRS is limited by particle trapping and wavebreaking \cite{Andreev2006} and can therefore support pump amplitudes several orders of magnitude higher than SRS \cite{Lancia2016signatures}.
Additionally, we make the argument
(9) that SBS may be appropriate in regimes where Langmuir waves would be collisionally damped. 

The relative advantages of Raman amplification are that higher peak intensities and amplification ratios may be achieved, and that the faster plasma-wave timescale allows compression to shorter pulse durations \cite{Trines2014}. Currently, experimentally demonstrated performance of Raman amplifiers \cite{Ren2008} exceeds that of Brillouin amplifiers \cite{Lancia2010,Lancia2016signatures} in energy transfer efficiency and peak amplified seed intensity, though higher total energy transfer between pump and seed has been achieved with SBS \cite{Lancia2016signatures}.

The possible advantages of Brillouin amplification described above can be roughly divided into several categories, based on whether they depend on the plasma-wave frequency [Re($\omega$)] (1-6), the linear growth rate [Im($\omega$)] (7), or the kinetic or non-linear properties of the Langmuir wave (8,9). The properties related to Re($\omega$) and Im($\omega$) may be examined to first order using the dispersion relation derived in Sec.~\ref{sec:disp}. In the following sections we will discuss the differences that arise from the instability frequency (1-6) and growth rate (7), including Landau damping (8). Collisional damping (9) will be discussed in Sec.~\ref{sec:collisional}.

\subsection{Brillouin Scattering Frequency}

\begin{figure*}
\centering
\includegraphics[width=\linewidth]{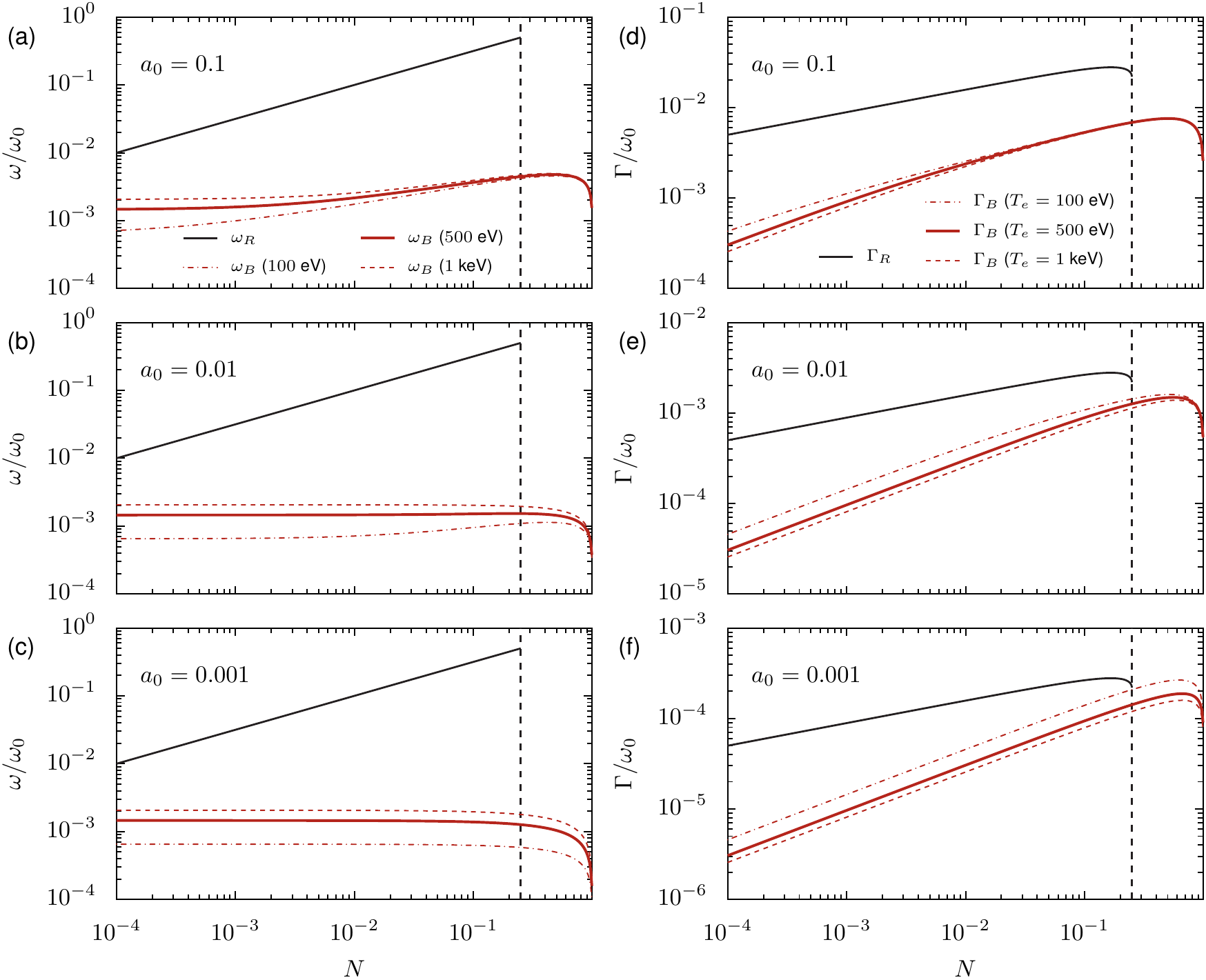}
\caption{(a-c) The real part of plasma wave frequency for both Raman ($\omega_R$) and Brillouin ($\omega_B$) scattering at varied $N$ and $a_0 = 0.1$ (a), $0.01$ (b), and $0.001$ (c), and $T_e = 100$, $500$, and $1000$ eV. $m_i/m_e = 1836$. (d-f) The undamped linear growth rate of Raman ($\Gamma_R$) and Brillouin ($\Gamma_B$) scattering under same conditions as the real-component calculations. The growth rate of Raman scattering is only non-zero for $N < 0.25$, and the growth rate of Brillouin scattering goes to zero for $N \ge 1$. The Brillouin growth rates and frequencies at different electron temperatures converge in the strong-coupling regime.}
\label{fig:Fixeda0}
\end{figure*}

The ion-wave frequency [$\textrm{Re}(\omega)$] is much smaller than the frequency of Langmuir waves, even in the strongly-coupled regime, since $m_i \gg m_e$, as illustrated by the calculated frequencies for both the Raman and Brillouin modes in Fig.~\ref{fig:Fixeda0}a-c. Note that the inequality $\omega_R  > \omega_B$ holds for all values of $a_0$ and $N$ in any reasonable domain. The primary advantages of Brillouin amplification come from this property, but it should be noted that this is at the direct cost of longer minimum compressed pulse durations, since Brillouin and Raman amplification cannot compress a pulse to a duration shorter than approximately a period of the relevant wave [$\approx 1/\mathrm{Re}(\omega)$]. This does not exclude the possibility of further compression altogether, e.g. superradiant amplification \cite{Shvets1998} can produce pulses shorter than a plasma period, but the governing equations we use here will no longer be valid and any further amplification should not be termed SRS or SBS. 

(1) The most important implication of the smaller acoustic frequency is that the pump and seed laser frequencies ($\omega_\textrm{pump} - \omega_{\textrm{seed}} = \omega_\textrm{plasma}$) may be almost the same for Brillouin amplification ($\omega/\omega_0 \approx 0.001$), whereas most Raman amplification studies are conducted with $\omega/\omega_0$ between 0.05 and 0.2. The limited availability of wavelengths for high-intensity short-pulse lasers restricts viable frequency separations, so support for a pump and seed at the same frequency is a useful attribute of SBS. 

(2) Since the frequency matching condition is fundamentally a statement of conservation of energy, i.e. a pump photon decays into a plasmon (or phonon) and a seed photon, the Manley-Rowe relations dictate that the maximum energy fraction transferred to the seed is $1 - \textrm{Re}(\omega)/\omega_0$. Inspection of Fig.~\ref{fig:Fixeda0}a-c suggests maximum Brillouin efficiencies of ~99.9\%, whereas the densities used for Raman amplification often give 80-90\% maximum energy transfer efficiency. Therefore, it is generally true that the maximum efficiency of SBS is higher, though the difference between 90\% and 99.9\% is likely to be small compared to other factors governing the amplification efficiency; the Manley-Rowe limit is difficult to reach. As an example, in the Raman wavebreaking regime the maximum level of pump depletion will not be achieved \cite{Edwards2015}, with Raman energy transfer efficiencies dropping to 10\% or lower, so in general the Manley-Rowe limit alone does not provide a solid justification for choosing a particular mode.

(3) SBS is considered more resilient to density fluctuations than SRS because the seed wavelength required to satisfy the frequency matching conditions does not depend substantially on the local plasma density; since the real frequency of SBS is small, the closely matched pump and seed frequencies always satisfy the matching condition. Other problems associated with plasma fluctuations, e.g. phase-front distortion of the seed or pump during propagation, equally affect SRS and SBS because they are not dependent on the frequency separation of the beams. Therefore, this additional robustness of SBS exists to the extent that density fluctuations disrupt the frequency matching conditions for Raman amplification. 

Plasma density fluctuations of 10\% lead to fluctuations in the local plasma frequency of around 5\%. For a plasma density $N = 0.01$, this corresponds to 1\% fluctuations in the resonant seed frequency. Since the FWHM frequency bandwidth of, for example, a 100 fs pulse is just over 2\% of the central frequency, density fluctuations of this magnitude on a scale longer than the plasma wavelength would moderately reduce but not significantly hinder the Raman amplification of ultra-short pulses. The finite allowable bandwidth of the plasma wave frequency provides additional resilience. The regime where plasma density fluctuations are substantial enough to make Raman amplification inferior to Brillouin amplification, yet not so large as to substantially disrupt the seed and pump propagation may therefore not be very large or significant. 

(4) Since the larger practical diameter of Brillouin amplification is primarily a result of increased resilience to plasma density fluctuations, the ability of SBS to support pulses
with higher total energy is true to the extent that (3) is. 

(5) Due to the frequency matching conditions, Raman amplification is not possible above $N = 0.25$, where $\omega_\textrm{seed} = \omega_\textrm{pump} - \omega_\textrm{plasma}$ gives $\omega_\textrm{seed} = \omega_\textrm{plasma}$, so in this regime, only Brillouin amplification is feasible. In Fig.~\ref{fig:Fixeda0} this is shown by the cutoff of the Raman lines above $N = 0.25$. This by itself is not necessarily an advantage, unless the plasma density is not at all controllable. Note that at $N>1$, the plasma is opaque to non-relativistic light, so neither process occurs. In the regime $0.25 < N < 1$, there is no thermally seeded SRS, removing this source of noise. However, thermally seeded SBS is still possible, so SBS in this regime has no real advantage over SRS at lower plasma densities with respect to spontaneous scattering seeded by noise.

\begin{figure}
\centering
\includegraphics[width=\linewidth]{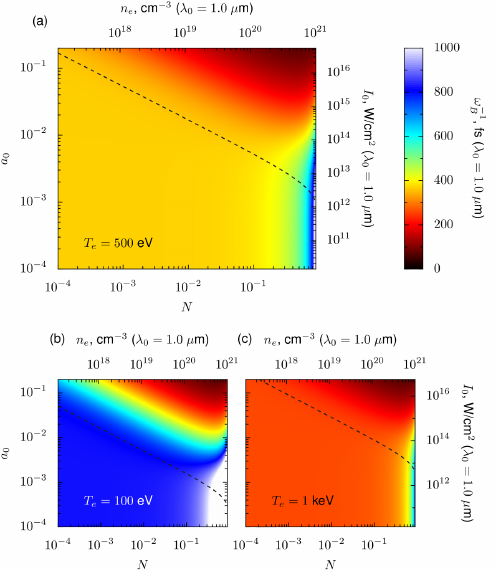}
\caption{The inverse of the real part of the Brillouin frequency ($\omega_B$) in femtoseconds (fs), which corresponds to the shortest pulse duration which can be produced via compression from Brillouin scattering. The dashed line marks the threshold for strongly-coupled Brillouin scattering ($\Lambda = 1/4$); the ion-acoustic frequency in the weakly-coupled regime, which lies below and to the left of the threshold, has no dependence on $a_0$ or $N$. The three plots correspond to electron temperatures of (a) 500 eV, (b) 100 eV and (c) 1 keV, and all calculations take $m_i/m_e = 1836$ and $Z = 1$. The scales presented in dimensional units assume a wavelength of 1 $\mu$m.}
\label{fig:BCompress}
\end{figure}

(6) A key motivation for the study of amplification by SC-SBS has been that it combines the above advantages with the ability to compress pulses to a duration within an order of magnitude of the minimum for Raman-compressed pulses \cite{Andreev2006}, based on a comparison between Brillouin amplification at $N = 0.3$ and Raman amplification at $N = 0.01$. However, at lower densities the real part of the SC-SBS frequency drops, increasing the minimum allowable pulse length. At three different temperatures, Fig.~\ref{fig:BCompress} shows the inverse real Brillouin frequency for pump and seed wavelengths at $\lambda_0 = 1$ $\mu$m as a function of $a_0$ and $N$, giving an approximate measure of the minimum compressed SBS pulse duration. Across reasonable conditions this ranges from 1 ps to about 100 fs, though the difference between the weakly- and strongly-coupled regimes is generally less than a factor of about 5. This limitation appears to rule out the generation of sub-100-fs pulses via Brillouin amplification, and severely restricts the usefulness of SBS in the sub-picosecond regime. Note that the plasma wave period is $2\pi/\omega_B$, i.e.~approximately six times longer than the inverse frequency, so the actual limit on pulse duration may be somewhat longer than the inverse frequencies in Fig.~\ref{fig:BCompress}. 

\subsection{Brillouin Scattering Growth Rate}

(7) The assertion that amplification by stimulated Brillouin scattering requires a plasma of shorter length than that necessary for SRS because the energy transfer is faster \cite{Andreev2006,Weber2013,Shoucri2015} demands some care. In the linear regime, the degree of amplification is set by the growth rate and the amplification distance, so this is fundamentally a statement about the respective growth rates of Raman and Brillouin scattering. In the self-similar (pump depletion) regime, the degree of achievable pump depletion determines how much energy is transferred; however, as discussed earlier, the difference in pump depletion between SRS and SBS is not sufficient to explain an orders-of-magnitude difference in plasma length. We will therefore consider the plasma length requirements that result from the linear growth rate. 

As originally formulated \cite{Andreev2006}, the comment on required plasma length was a comparison between SRS at $N \approx 0.01$ and SBS at $N = 0.3$, though it also has been repeated as justification for later studies, which considered SBS at $N < 0.1$. Since the Raman growth rate is higher than the Brillouin growth rate for physical ($m_i/m_e \ge 1836$) plasmas at every value of $N$ and $a_0$ where both modes can be present (see Fig. \ref{fig:GamBtoR}), switching from SRS to SBS at fixed $N$ and $a_0$ will not allow a shorter plasma for the same degree of amplification. Figure \ref{fig:Fixeda0}d-f allows us to consider the original statement, showing the linear Raman and Brillouin amplification growth rates at three distinct fixed pump intensities over a wide range of $N$. It is immediately apparent that the maximum Raman growth rate is generally much higher than the maximum Brillouin growth rate, and that only for $a_0 = 0.001$ does the Brillouin growth rate at $N = 0.3$ exceed the Raman growth rate at $N = 0.01$. This value of $a_0$ lies well within the weakly-coupled regime, nullifying the pulse-compression benefits of the strong-coupling regime. On the basis of the relative growth rates of SRS and SBS, it cannot be generally argued that the plasma length for a SC-SBS-based amplifier will be shorter than that for a Raman-based amplifier with the same degree of amplification.

(8) At sufficiently high electron temperatures Langmuir waves are Landau damped, reducing the efficiency of Raman amplification \cite{Hur2005,Malkin2007,Yampolsky2009,Malkin2010,Yampolsky2011,Strozzi2012,Wu2014,Depierreux2014}. Landau damping of the ion-acoustic wave is primarily sensitive to the ratio $Z T_i /T_e$, so for $Z T_i /T_e \ll 1$ and large $T_e$, only SRS will be suppressed. It must be noted, however, that significant amplification of pulses can be maintained with significant Landau damping in the quasi-transient backward Raman amplification (QBRA) regime \cite{Malkin2009quasitransient}, so the electron temperature required to suppress the Raman mode may be somewhat higher than would normally be anticipated \cite{Malkin2009quasitransient,Malkin2010,Balakin2011numerical}. For QBRA, as for other forms of Raman amplification, the center of the amplified seed spectrum is down-shifted from the pump frequency by the plasma frequency. Here we consider the electron temperature required to suppress Raman amplification below the Brillouin amplification growth rate across a broad range of conditions. 

\begin{figure}
\centering
\includegraphics[width=\linewidth]{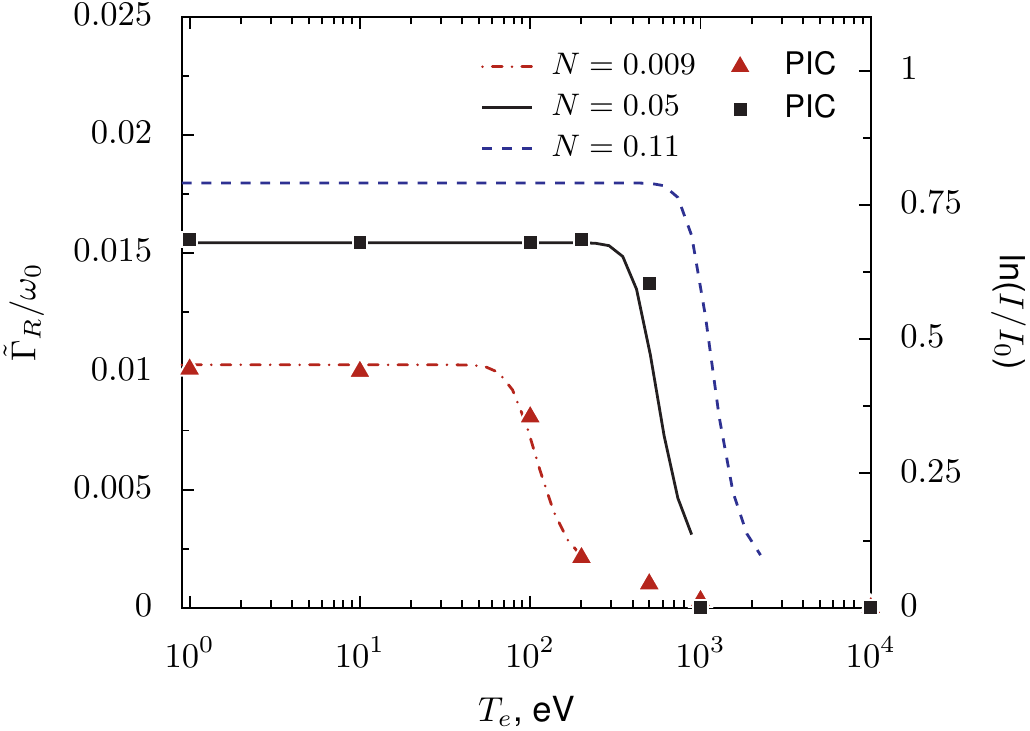}
\caption{Raman amplification growth rate, modified to account for Landau damping, as a function of electron temperature at varied $N$. The logarithm of the final to initial intensity ratios found in PIC simulations at $N = 0.009$ and $0.05$ and $a_0 = 0.067$ is plotted normalized to the growth rate at $T_e = 1$ eV.}
\label{fig:Landau1}
\end{figure}

Drawing on previous work \cite{Malkin2009quasitransient}, we can modify the expression for the Raman growth rate to take damping into account:
\begin{equation}
\tilde{\Gamma}_R = \frac{\Gamma_R}{\sqrt{1+\tilde{\nu}} + \tilde{\nu}}
\label{eqn:landamp1}
\end{equation}
where 
\begin{equation}
\tilde{\nu} = \frac{\nu}{2 \Gamma_R}
\end{equation}
and $\nu$ is the damping rate. $\Gamma_R$ is the unmodified growth rate in the linear regime, and $\tilde{\Gamma}_R$ is the rate at which the seed field strength peak is expected to grow in the Landau-damped regime. To calculate $\nu$ for Landau damping, we assume that the electron velocity distribution is thermal and note that Landau damping is proportional to the slope of the velocity distribution, giving \cite{Malkin2009quasitransient}:
\begin{equation}
\nu = \frac{\omega_e \sqrt{\pi}}{(2 q_T)^{3/2}} e^{\left(- \frac{1}{2q_T}\right)}
\label{eqn:landamp2}
\end{equation}
where $q_T = T_e/T_m$ and $T_m = mc^2 N /4$. This analysis is only valid for $T_e \ll T_m$, so our results have been truncated to exclude densities and temperatures which violate this condition. Figure \ref{fig:Landau1} shows the modified growth rates for Raman amplification as a function of electron temperature for different plasma densities. The analytic predictions from the above equations are compared to PIC simulations of Raman amplification. Both the theory and simulations show little dependence on temperature until a cutoff value of $T_e$ where the expected and observed growth rapidly decreases; this threshold depends on both plasma density and laser intensity. 

\begin{figure*}
\centering
\includegraphics[width=\linewidth]{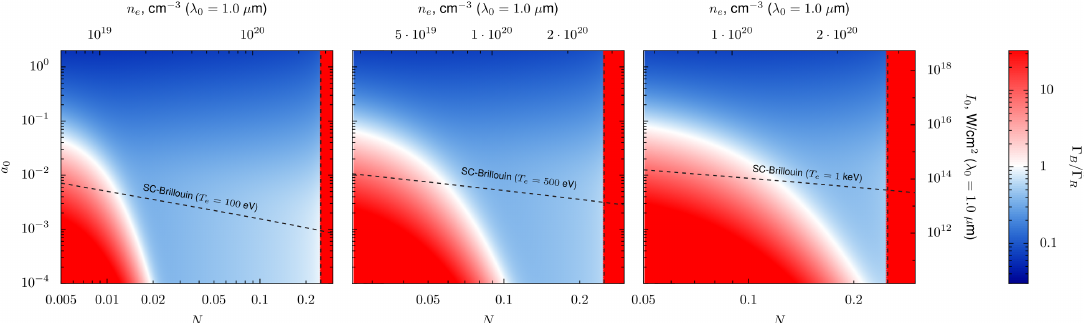}
\caption{Ratio of the Brillouin growth rate to the Raman growth rate with the correction for Landau damping of the Langmuir wave at $T_e = 100$, $500$, and $1000$ eV. The ion temperature is assumed to be negligible, so that Landau damping of the acoustic wave is not significant. The strong-coupling threshold is marked for each temperature. These plots suggest that Landau damping only opens a very narrow window where $\Gamma_B > \tilde{\Gamma}_R$ in the strongly-coupled Brillouin scattering regime. The above plots are based on Eqs.~\ref{eqn:landamp1}-\ref{eqn:landamp2}. Note that the x-axis of each plot is different because the analysis is only valid for $T_e \ll T_M$ and the axis has been modified to exclude regions where the model is not valid.}
\label{fig:Landau2}
\end{figure*}

If we compare the modified Raman growth rate to the Brillouin growth rate at varied $a_0$, $N$, and $T_e$, as in Fig.~\ref{fig:Landau2}, we can identify regions where the modified Raman growth rate is smaller than that for the Brillouin mode. The above simple equations predict that only a very narrow region exists at any temperature where $\Gamma_B$ is larger than $\tilde{\Gamma}_R$ above the SC-SBS threshold, so little flexibility is available for using Landau damping to preferentially amplify by SC-SBS. 

The above analysis fails in regimes where the velocity distribution of electrons is strongly non-thermal; in particular, the wavebreaking and superradiant regimes may be more robust to damping than suggested by this formulation. However, over a broad range of important parameters, the model provides a reasonable estimate of Landau damping effects, showing agreement for predicted temperature cutoffs with PIC simulations. 

Landau damping may be a useful mechanism for suppressing Raman amplification so that Brillouin amplification can be studied either computationally or experimentally. However, when considering the design of plasma amplifiers, the necessity of suppressing amplification by SRS is this fashion for a broad range of regimes may suggest that SRS would be a more suitable process for pulse amplification. 

It has also been suggested that in the Raman wavebreaking regime (the threshold of which is shown in Fig.~\ref{fig:GamBtoR}), the lower efficiency of Raman amplification may make Brillouin amplification more promising. This is a claim that must be treated with care because the mechanism of suppressed Raman amplification in this regime is somewhat different than that of Landau or collisional damping \cite{Edwards2015,Farmer2015raman}. In particular, although the efficiency of Raman amplification drops in the wavebreaking regime, the growth rate of the pulse front is still high and the plasma phase space after the pulse front is substantially disrupted, hindering any subsequent amplification by SBS. Furthermore, for sufficiently short and intense seeds, we may enter the superradiant regime \cite{Shvets1998}, where substantial pump depletion may occur at the leading edge of the pulse, suppressing subsequent Brillouin growth. 

\section{Limits of Brillouin Amplification}
\label{sec:lims}
Stimulated Raman backscattering is not the only instability which may limit the utility of Brillouin amplification. In this section we quantify the problems posed by the forward Raman scattering, filamentation, and the resonance condition. 

\subsection{Forward Raman Scattering}

In sub-quarter-critical density plasmas stimulated Raman backscattering is  suppressed at increased electron temperatures due to Landau damping. However, as has been suggested previously \cite{Alves2013,Trines2014}, forward Raman scattering of the \emph{seed} may still interfere with Brillouin amplification. The resonance conditions for forward Raman scattering of the seed are: $\omega_{\rm seed} = \omega_{\rm plasma} + \omega_{\rm scattered}$ and $k_{\rm seed} = k_{\rm plasma} + k_{\rm scattered}$, so that the scattered light is downshifted from the seed by the plasma frequency. However, since the seed and the scattered light copropagate, $k_{\rm plasma}$ must now be the difference between $k_{\rm seed}$ and $k_{\rm scattered}$, i.e.:
\begin{equation}
k_{\rm plasma} = \frac{\omega_0}{c} \left[1 - \left(1 - \frac{\omega_e}{\omega_0}\right)\right] = \frac{\omega_e}{c}
\end{equation}
The plasma wave in forward Raman scattering has a phase velocity of $c$ and is therefore not readily Landau damped. The growth rate of the instability is given by \cite{Trines2014,Gibbon2005}:
\begin{equation}
\label{eqn:frs}
\Gamma_{RFS} = \frac{b_0}{2 \sqrt{2}} \frac{\omega_e^2}{\omega_0}
\end{equation}
where $b_0 = E_{0,\textrm{seed}}/E_\textrm{rel}$ is the normalized seed amplitude. We focus on the seed because at fixed $a_0$ and standard conditions for stimulated Brillouin scattering, the RFS growth rate is less than that of SBS, so RFS of the pump should not pose as serious an issue. 

The ratio of seed field strength ($b_0$) to pump field strength ($a_0$) at which the growth rate of RFS from the seed equals the growth rate of Brillouin backscattering, i.e. $\Gamma_{RFS} = \Gamma_{SBS}$, marks an approximate threshold for the regime where RFS must be considered. In the strongly-coupled regime this may be written analytically from Eqs.~\ref{eqn:scb} and \ref{eqn:frs}, noting that $2^{1/6}3^{1/2} \approx 1.94$:
\begin{equation}
\frac{b_0}{a_0} = 1.94 \left(\frac{Z m_e}{m_i}\right)^{\frac{1}{3}} \frac{1}{a_0^{\frac{1}{3}} N^{\frac{2}{3}}}
\end{equation}

\begin{figure}
\centering
\includegraphics[width=\linewidth]{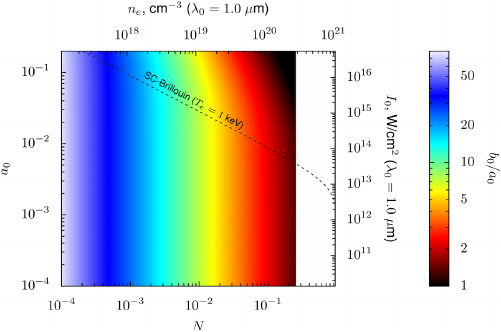}
\caption{Ratio of seed field strength ($b_0$) to pump field strength ($a_0$) at which the growth rates of RFS from the seed and SBS from the pump are equal. The dashed line marks the strong-coupling threshold. This plot is calculated for $T_e = 1$ keV and $m_i/m_e$ = 1836. The result of the seed strength greatly exceeding the presented thresholds is strong envelope modulation caused by the presence of both the fundamental and plasma-downshifted frequencies, leading to breakup of the seed pulse. }
\label{fig:RFS1}
\end{figure}

In Fig.~\ref{fig:RFS1} the seed to pump field strength ratio at which the growth rates are equal is plotted for $T_e = 1$ keV and $m_i/m_e = 1836$. Though this should not be viewed as an exact 
threshold (the growth rate is not the same as the rate at which energy is actually being shifted between modes), the ratio $b_0/a_0$ for which $\Gamma_{RFS} = \Gamma_{SBS}$ approximates the maximum seed intensity which can be reached before RFS begins to detrimentally affect seed development. Of particular concern for Brillouin amplification is that for $N>0.05$, $b_0/a_0$ is less than 3, so it may not be possible to amplify a seed to an intensity more than 10 times that of the pump for $0.05 < N < 0.25$. Since the goal of parametric plasma amplification is to generate pulses of higher intensity than what can be produced in conventional solid-state systems, e.g. the system providing the pump beam, this limit imposes a severe constraint on the usefulness of SBS in this regime. However, it should be noted that RFS, like Raman backscattering, is not a problem for $N >0.25$.

\begin{figure}
\centering
\includegraphics[width=\linewidth]{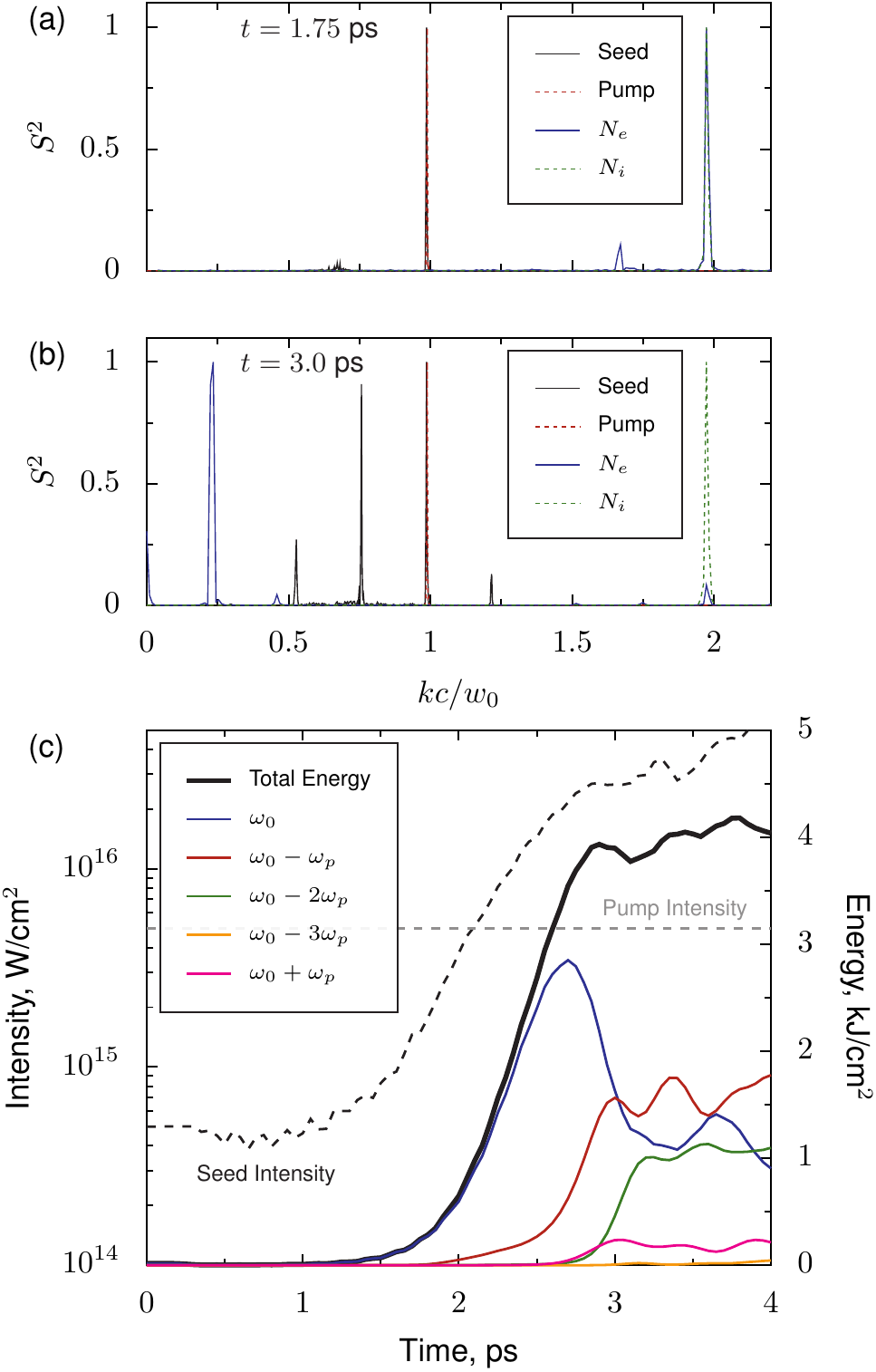}
\caption{PIC simulation of Brillouin amplification at $N= 0.05$ where stimulated Raman backscattering is heavily Landau damped ($T_e = 5$ keV, $m_i/m_e = 1836$). (a,b) Spectra of seed laser, pump laser, electron density fluctuations ($N_e$) and ion density fluctuations ($N_i$) before (a) and after (b) onset of significant forward Raman scattering, showing appearance of electron density fluctuations at $k = \omega_e/c$. Each spectrum is normalized to its maximum. (c) Intensity of seed and energy distribution in different seed frequencies as a function of amplification time. The Raman forward scattering growth rate equals the Brillouin backscattered growth rate at $t \approx 2.5$ ps when $I_\textrm{seed} = 4 I_\textrm{pump}$ ($b_0/a_0 = 2$). The pump intensity is $5 \times 10^{15}$ W/cm$^2$ ($a_0 = 0.06$), the ion temperature is $T_i = 10$ eV and the mass ratio is $m_i/m_e = 1836$ with $Z = 1$. The simulations used 80 cells/$\lambda_0$ and 150 particles per cell.}
\label{fig:RFS2}
\end{figure}

Figure \ref{fig:RFS2} shows a simulation of plasma amplification with heavily suppressed SRS ($T_e = 5$ keV). The amplification is attributable to SBS, as confirmed by the narrow spectrum of the amplified seed and the characteristic matching of the ion and electron density fluctuations. However, where the ratio $b_0/a_0$ exceeds 2, which for $T_e = 5$ keV, $m_i/m_e = 1836$, $a_0 = 0.06$ ($I = 5\times10^{15}$ W/cm$^2$), and $N = 0.05$ is where $\Gamma_{RFS} = \Gamma_{SBS}$, a significant fraction of the seed energy begins to shift to the frequency $\omega_0 - \omega_p$, corresponding to the Stokes component of RFS. On the basis of the frequency composition, this could be the result of Raman backscattering, but strong fluctuations in the electron density at $k = \omega_{pe}/c$ are characteristic of RFS, as seen in Fig.~\ref{fig:RFS2}b, and at this later time there are no significant fluctuations at $k = 2k_0 - \omega_{pe/c}$. The plot of seed intensity in Fig.~\ref{fig:RFS2} shows that the maximum seed intensity does not increase much beyond the level where the two growth rates are equal. 

Light shifted away from $\omega_0$ will no longer contribute to the amplification process, since the backward Raman mode is suppressed and only backward Brillouin scattering can transfer energy from the pump. Strong RFS eventually halts pulse compression and amplification. The more pressing problem is that the presence of significant energy at the downshifted frequency co-propagating with the seed causes envelope modulations at the frequency difference. These modulations strongly interact with the plasma and eventually cause the breakup of the seed pulse. Though the noise that seeds RFS is overestimated in PIC simulations, i.e. RFS may not be quite so catastrophic for experiments, the high growth rate of RFS relative to SBS means that smaller levels of noise will only delay, not suppress, adverse RFS effects. 

RFS does not affect Raman amplification as seriously as Brillouin amplification because the RFS growth rate is lower than the RBS growth rate and SRS amplification  is usually implemented at lower plasma densities than SBS amplification. When $\Gamma_{RBS} = \Gamma_{RFS}$:
\begin{equation}
\frac{b_0}{a_0} = \frac{\sqrt{2}}{N^{3/4}}
\end{equation}
which has a minimum value of $4$ at $N = 0.25$ and increases to $45$ at $N = 0.01$. 

\subsection{Filamentation}

Amplification may also be limited by filamentation, or the transverse collapse of a laser beam, which can be caused by ponderomotive \cite{Kruer2003,Kaw1973filamentation}, relativistic \cite{Max1974self,Decker1996evolution}, or thermal \cite{Perkins1974thermal,Epperlein1990kinetic} changes in the plasma index of refraction. The fundamental mechanism of filamentation is an intensity-dependent change in the index of refraction causing self-focusing of a beam. In plasmas, this may result from reduction in the plasma density by heating or ponderomotive expulsion of electrons and ions, or a reduction in the plasma frequency due to relativistic mass increase of electrons in intense fields. Difficulty associated with filamentation instabilities at high plasma densities has been a key motivation for studying the utility of SBS at $N < 0.25$ \cite{Weber2013,Trines2014}.

To describe the growth of ponderomotive filamentation, we can write a dispersion relation describing transverse instability ($\mathbf{k} \cdot \mathbf{k}_0 = 0$) \cite{Kruer2003}:
\begin{equation}
\left(\Gamma_{PF}^2 + k^2 c_s^2\right) \left(\Gamma_{PF}^2 + \frac{k^4c^4}{4 \omega_0^2} \right) = \frac{k^4 v_{osc}^2 c^2}{8} \frac{\omega_{pi}^2}{\omega_0^2}
\label{eqn:filpf}
\end{equation}
which for $\Gamma_{PF} \ll kc_s$ becomes \cite{Kruer2003}:
\begin{equation}
\Gamma_{PF} = \frac{1}{8} \left(\frac{a_0 c}{v_e}\right)^2 \frac{\omega_{pe}^2}{\omega_0}
\end{equation}
and for $\Gamma_{PF} \gg kc_s$ is \cite{Max1974self}: 
\begin{equation}
\Gamma_{PF} = \frac{a_0 \omega_{pi}}{\sqrt{2}}
\end{equation}
Since $\Gamma_{PF}$ is comparable to $kc_s$ for many parameters of interest in SBS, Eq.~\ref{eqn:filpf} may be evaluated numerically, solving for the value of $k$ which produces the maximum value of $\Gamma_{PF}$. 

\begin{figure}
\centering
\includegraphics[width=\linewidth]{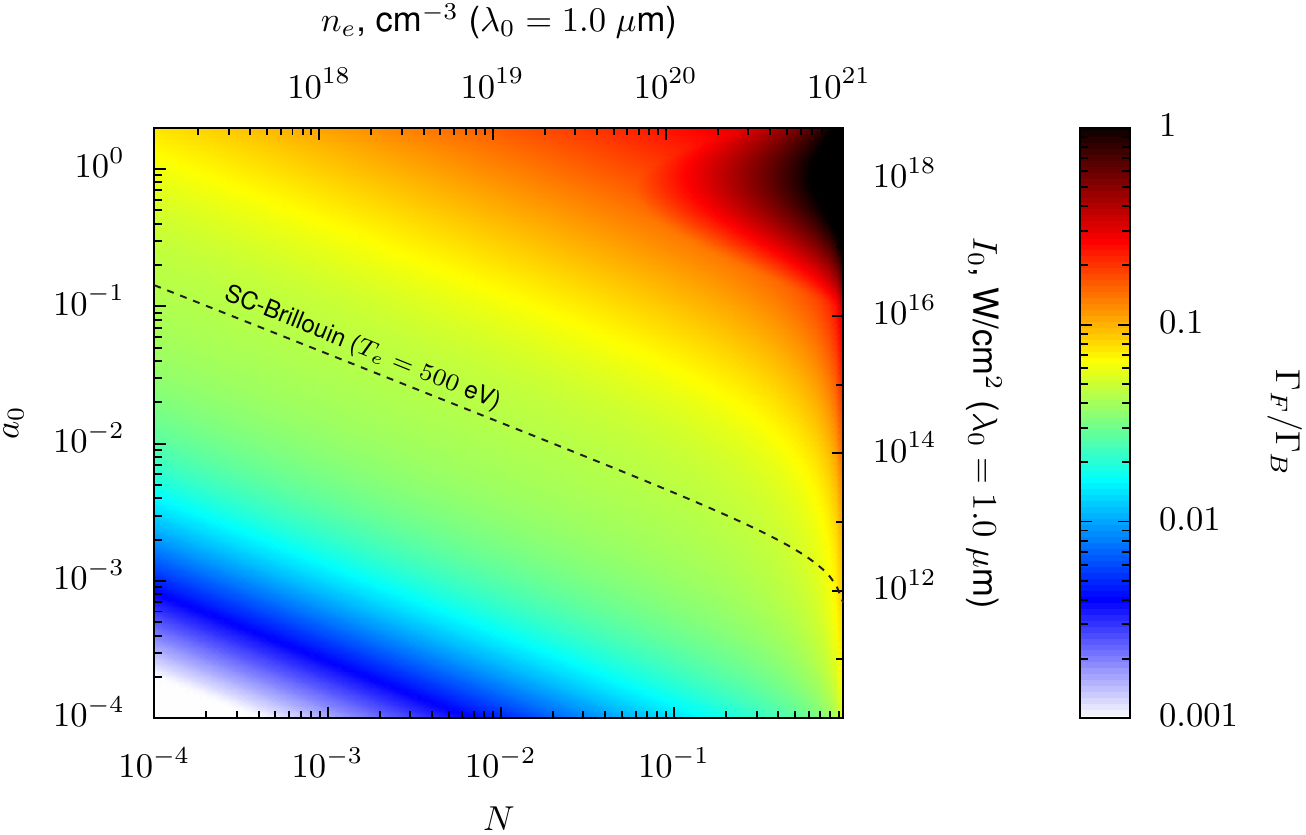}
\caption{The ratio of the maximum filamentation growth rate to the Brillouin growth rate as a function of both $a_0$ and $N$ for $T_e = 500$ eV and $m_i/m_e = 3600$. The ponderomotive filamentation rate is higher than the relativistic filamentation rate, apart from approximately where $\Gamma_F/\Gamma_B >0.2$. The strongly-coupled Brillouin scattering threshold is marked as a reference.}
\label{fig:Fil1}
\end{figure}

Figure \ref{fig:Fil1}, shows the maximum filamentation growth rate compared to the growth rate of Brillouin backscattering. Ponderomotive filamention, as calculated from Eq.~\ref{eqn:filpf}, is the dominant mechanism for most parameters of interest, apart from high intensity lasers in dense plasma, where relativistic filamentation is more significant. The growth rate for relativistic filamentation \cite{Decker1996evolution} is given by:
\begin{equation}
\Gamma_{RF} = \frac{1}{8} \frac{a_0^2}{\left(1+a_0^2\right)^{3/2}} \frac{\omega_{pe}^2}{\omega_0}
\end{equation}
In Fig.~\ref{fig:Fil1} relativistic filamentation is the more important mechanism approximately where $\Gamma_F/\Gamma_B > 0.2$. For short pulses in regimes that are not strongly collisional, heating is too slow to substantially affect the pulse and thermal filamentation does not play a strong role.

As shown by Fig.~\ref{fig:Fil1}, the relative growth rate of filamentation increases at higher plasma densities, so decreasing plasma density at fixed pump intensity will decrease the role of filamentation. Figure \ref{fig:Fil1} compares the growth rate of pump filamentation to the Brillouin scattering growth rate of the seed, but for high seed intensities, filamentation of the seed may cause a more rapid disruption of the amplification process. The role of seed filamentation in high intensity regimes can be estimated by noting that $\Gamma_{PF}$ increases linearly with $a_0$ (for $\Gamma_{PF} \gg kc_s$).  

\subsection{Resonance Conditions}
Though it is sometimes assumed that the sufficient condition for driving SBS instead of SRS is that the frequency difference between the pump and the seed corresponds to the ion-acoustic frequency, in practice the situation can be somewhat more complicated \cite{Riconda2013,Jia2016}, especially for ultra-short pulses. In particular, though it may appear that the plasma frequencies required for Raman and Brillouin amplification are well-separated, the exponential nature of the instability can radically and rapidly change the frequency distribution of the system. Application of the resonance conditions therefore requires some care.

\begin{figure}
\centering
\includegraphics[width=\linewidth]{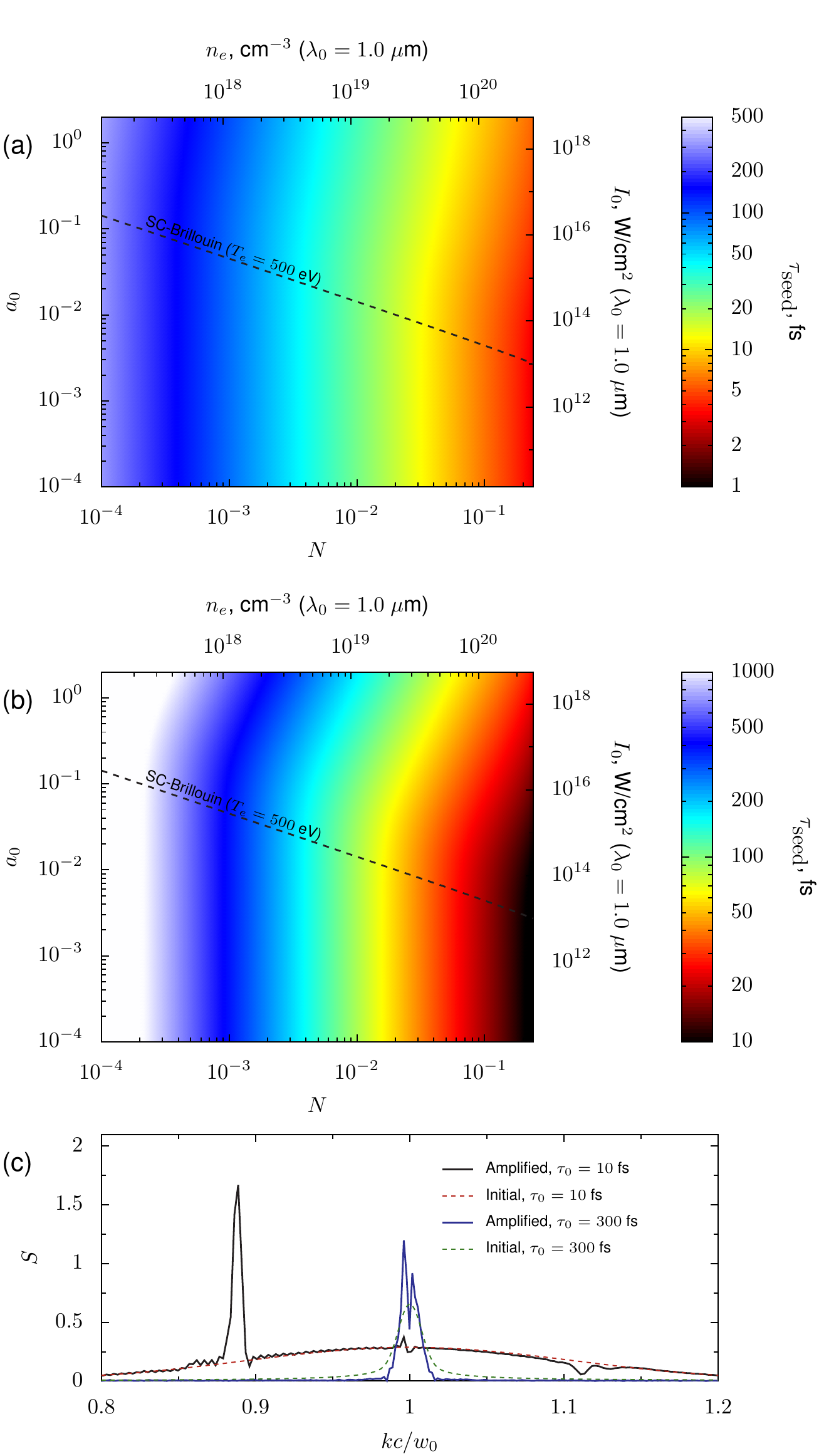}
\caption{(a) Under the assumption that the frequency components of the seed pulse will be independently amplified at their respective growth rates, this plot gives the incident seed pulse length for which an initially Brillouin-centered pulse will instantaneously have equal Raman and Brillouin growth rates simply due to the frequency spread of the initial pulse at $T_e = 500$ eV and $m_i/m_e = 3600$. (b) The seed duration at which Raman and Brillouin amplification will equally contributed if 100-fold Brillouin amplification is attempted. This represents the minimum pulse duration for which Brillouin amplification is possible. $T_e = 500$ eV and $m_i/m_e = 3600$. (c) At $N = 0.01$ and $a_0 = 0.02$, the initial and amplified spectra of seeds with 10 and 300 fs FWHM durations found with PIC simulations, showing the preferential Raman amplification that occurs due to the initial spectral spread of short pulses. $T_e = 1$ keV, resulting in a slight downshift of the Raman peak from the cold plasma frequency. Although the 10 fs initial seed is short enough to satisfy the superradiant criteria, the simulation lies below the superradiant threshold and the amplified pulse has a longer duration.}
\label{fig:Res1}
\end{figure}

The frequency and wavenumber spread of each of the three waves means that a finite bandwidth may satisfy the three-wave coupling conditions. Given a growth rate ratio $\Gamma_B/\Gamma_R$ (generally $\ll 1$) and an ultrashort Gaussian seed pulse, a limit can be set on how short an ideal pulse may be before its spectral width is sufficient to drive a significant portion of transferred energy into the Raman mode. The below analysis can be regarded as a best-case for Brillouin amplification, since for both experimental and computational pulses, spectra tend to be wider than a Fourier-transform-limited Gaussian. The change in intensity due to a particular mode $I_m$ with instability growth rate $\Gamma_m$ is given by:
\begin{equation}
I_m = I_{m,0} e^{2 \Gamma_m t}
\end{equation}
where $I_{m,0}$ is the initial intensity of the seed frequency component corresponding to that mode. If we consider amplification in the linear regime, we can work out which mode will contribute more to the final, amplified pulse, as:
\begin{equation}
\frac{I_B}{I_R} = \frac{I_{B,0} e^{2 \Gamma_B t}}{I_{R,0} e^{2 \Gamma_R t}} = \frac{I_{B,0}}{I_{R,0}} e^{2 \Gamma_B t (1 - \Gamma_R/\Gamma_B)}
\end{equation}
where $I_B$ represents the intensity at the Brillouin frequency, and $I_R$ represents the intensity considering only the component at the Raman frequency. Instantaneously ($t = 0$), the relative rate of increase in intensity for each mode is:
\begin{equation}
\frac{\partial_t I_B}{\partial_t I_R} = \frac{I_B}{I_R} \frac{\Gamma_B}{\Gamma_R}
\end{equation}
For an initial Gaussian pulse centered at the the pump wavelength, the ratio $I_{B,0}/I_{R,0}$ can be worked out from the pulse duration, noting that the pulse full-width-half maximum (FWHM) in the time and frequency domains are related by:
\begin{equation}
W_\omega = \frac{8 \ln 2}{W_t}
\end{equation}
where $W_t$ is the intensity FWHM duration in time and $W_\omega$ is the FWHM in frequency. For a pulse centered at the pump frequency:
\begin{equation}
\ln \frac{I_{B,0}}{I_{R,0}} = \frac{(\omega_e W_t)^2}{16 \ln 2}
\end{equation}
because the frequency separation between the two modes is approximately $\omega_e$. The seed duration for which a seed centered at the Brillouin shifted frequency will by amplified at equal rates by SRS and SBS due to its own natural bandwidth is plotted in Fig.~\ref{fig:Res1}a over a range of $a_0$ and $N$. The instantaneous growth rate is not as limiting as the ion-wave period, since the seed duration limits here are shorter than the SBS period, but spectral width becomes more of a problem if a substantial degree of amplification is desired.

As an example, consider an attempt to amplify a pulse by a factor of 100 using the Brillouin mode ($I_B = 100 I_{B,0}$). This sets $t$, and we can find the threshold incident pulse duration for which the final pulse contains equal energy at the Brillouin and Raman shifted frequencies using an expression for arbitrary amplification factor:
\begin{equation}
\left(W_t \omega_0\right)^2 = \frac{16 \ln 2}{N}  \left(\frac{\Gamma_R}{\Gamma_B} -1\right) \ln \left(\frac{I_B}{I_{B,0}}\right)
\end{equation}
For 100-fold amplification, this becomes:
\begin{equation}
W_t \omega_0  \approx \sqrt{\frac{51}{N} \left(\frac{\Gamma_R}{\Gamma_B} -1\right)}
\end{equation}
This limit is plotted in Fig.~\ref{fig:Res1}b for varied $N$ and $a_0$ using calculated values for $\Gamma_R$ and $\Gamma_B$ and assuming a wavelength of $1$ $\mu$m. For both experiments and PIC simulations the picture is somewhat less promising for Brillouin amplification, since real and simulated pulses are not as close to the Fourier transform limit as an ideal Gaussian, resulting in more initial light at the Raman shifted frequency. The spectral width of sub-picosecond seed pulses may therefore be problematic. In Fig.~\ref{fig:Res1}c, the spectra of initial and amplified seed pulses with initial durations of 10 and 300 fs are shown, demonstrating that for the same conditions, the seed duration alone can dictate whether SRS or SBS is the primary amplification mechanism.

\section{Brillouin Amplification in a Collisional Regime}
\label{sec:collisional}
From the above analysis, in a homogeneous collisionless plasma Raman amplification will generally be preferable to Brillouin amplification, but there are regimes where suppression of SRS may make Brillouin amplification the only viable route forward. Here we discuss a regime where SBS becomes the preferred method of amplification due to collisional damping of SRS. 

\begin{figure}
\centering
\includegraphics[width=\linewidth]{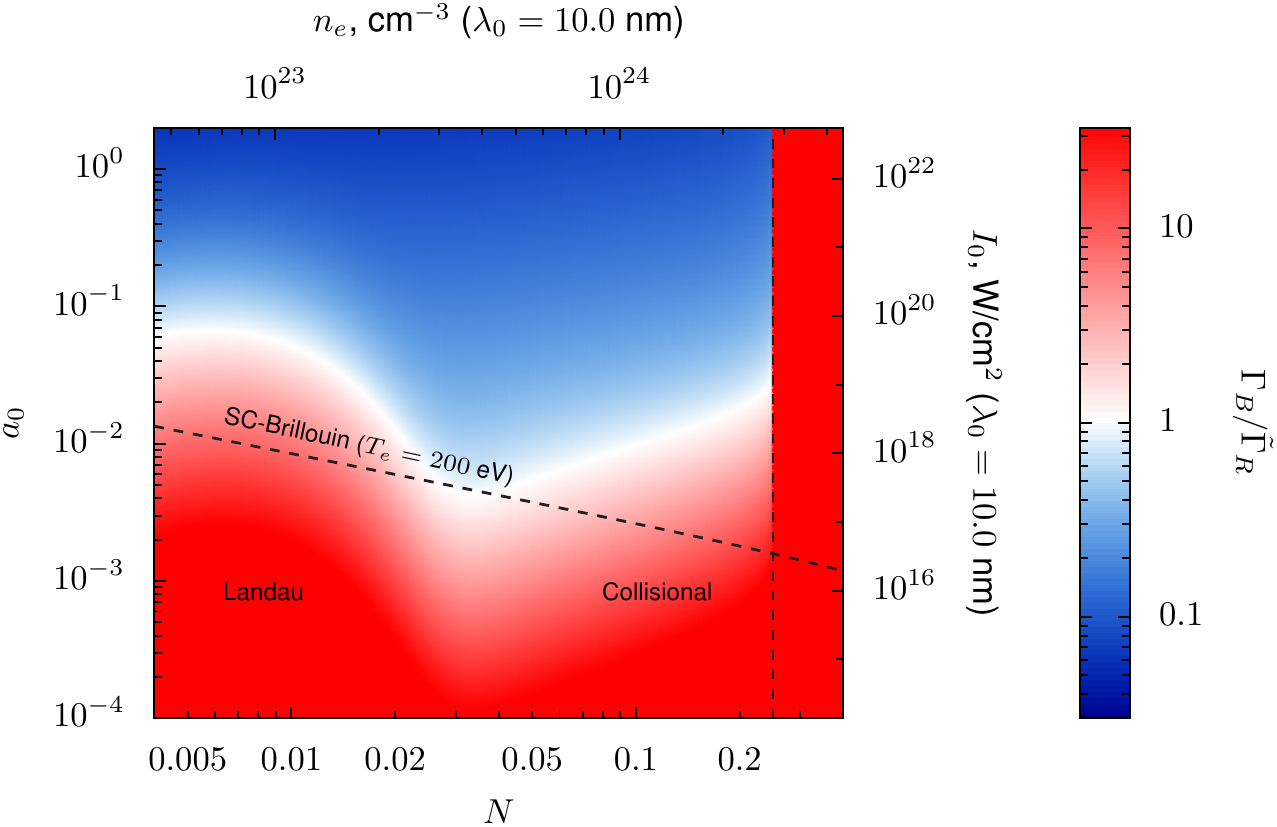}
\caption{Ratio of Brillouin growth rate to Raman growth rate with Landau and collisional damping of Raman growth calculated. The calculation assumes a wavelength of $\lambda_0 = 10$ nm for the pump laser, an electron temperature of $200$ eV, and a negligible ion temperature. The regions where Brillouin amplification shows a larger growth rate are labeled based on whether collisional or Landau damping suppresses the Langmuir wave. At higher temperatures and longer wavelengths collisional damping is less significant.}
\label{fig:Collisional1}
\end{figure}

Brillouin amplification in the presence of collisions \cite{Humphrey2013} and the effect of collisional damping on Raman amplification \cite{Malkin2009quasitransient} have both been considered previously. It has been suggested that Brillouin amplification is somewhat improved with collisions \cite{Humphrey2013}, and noted that as wavelength decreases, the window for which SRS is not significantly suppressed by Landau or collisional damping gradually closes, setting a lower limit on the wavelengths of light that can be amplified by SRS \cite{Malkin2007,Malkin2007compression,Malkin2009quasitransient,Malkin2010}. To compare Raman and Brillouin amplification in the presence of collisions, we may use a previously analyzed expression for the collisional damping rate \cite{Malkin2009quasitransient}:
\begin{equation}
\nu_{cln} = \frac{2 \sqrt{2}}{3 \sqrt{\pi}} \frac{\Lambda r_e \omega_0^3}{q_T^{3/2} c \omega_e}
\end{equation}
where $r_e = e^2/mc^2 \approx 2.818 \times 10^{-15}$ m and $\Lambda$ is the Coulomb logarithm. In Fig.~\ref{fig:Collisional1}, the Brillouin growth rate is compared to the Raman growth rate corrected for collisional and Landau damping for a short wavelength ($\lambda_0 = 10$ nm) pump at $T_e = 200$ eV. Under these interesting, albeit extreme, conditions, collisional damping plays a significant role, and over a broad range of densities and pump intensities, Brillouin amplification provides higher growth rates, including into the strongly-coupled regime. If collisional damping is strong, collisional absorption of the seed and pump pulses will limit the practicality of both SRS and SBS, but because SBS is not Landau damped under the same conditions as SRS, an SBS amplifier has more room in density-temperature space to avoid detrimental collisional effects. Amplification of visible light by both SRS and SBS is less affected by collisions due to the lower plasma densities required.  

As an example of the collisional regime, consider a pump beam at 10 nm with an intensity of $3.8 \times 10^{17}$ W/cm$^2$ ($a_0 = 0.005$), which is approximately the 3 GW peak power of the free-electron laser FLASH \cite{Schreiber2015free} focused to a 10 $\mu$m diameter spot. In a plasma with density $2 \times 10^{24}$ cm$^{-3}$ ($N = 0.18$), negligible ion temperature and 200 eV electron temperature, the growth rate for SBS is $\Gamma_B = 1.4 \times 10^{14}$ s$^{-1}$. In comparison, due to collisional damping, the above equation predicts an effective Raman growth rate of $4\times10^{13}$ s$^{-1}$. SBS therefore offers a significant advantage over Raman amplification in the short-wavelength, high-density regime associated with the amplification of x-ray pulses.

\section{Conclusion}
\label{sec:conc}

In this paper we examine parametric plasma amplification of sub-picosecond pulses by stimulated Brillouin backscattering. Though SBS is a mechanism that has been repeatedly considered theoretically, computationally, and, increasingly, experimentally, a systematic consideration of the useful regimes for Brillouin amplification based on the accepted mechanism models has thus far been lacking. We show that there are a number of regimes and applications for which plasma-based Brillouin amplification is both an interesting and promising route towards the next generation of high-intensity lasers, but also point out that the benefits of SBS are not as broad as often claimed, and many signals which have been attributed to stimulated Brillouin scattering are likely to be stimulated Raman scattering or electron-based superradiance, especially when considering the amplification of ultra-short pulses. 

Specifically, the SBS amplification mechanism allows the pump and seed lasers to be at the same frequency, which may make implementation of high-intensity systems more practical. However, the maximum efficiency of SBS is not substantially higher than SRS, nor is the increased robustness of SBS to density inhomogeneities likely to be the dominant factor under most conditions. Furthermore, because the growth rate of SBS is much lower than that of SRS, the plasma component of an SBS amplifier must be longer than that of an SRS amplifier to achieve the same degree of seed pulse amplification for the same pump intensity. SC-SBS ameliorates some of the problems associated with the long ion-acoustic wave period in weakly-coupled Brillouin scattering, but in general even SC-SBS is not directly competitive with SRS in terms of amplification factors and achievable pulse intensities. Furthermore, some care must be taken with the threshold for the strong-coupling regime and the conditions under which the strongly-coupled equations may be used. 

In general, seed pulses amplified by SBS will be less intense and longer in duration than those that can be achieved with SRS. For some applications this will be acceptable, and under conditions where suppression of SRS is unavoidable, e.g. short-wavelength collisional regimes, SBS may be a viable and useful option. We suggest, in fact, that because of strong collisional and Landau damping of SRS in dense plasmas at x-ray wavelengths, SBS may be the only option for amplification. However, in regimes where amplification by SBS can only be achieved via artificial suppression of SRS, e.g. through Landau damping or by the introduction of density gradients, SRS is likely to be the better choice for producing high-intensity pulses. Although SBS is more robust to plasma inhomogeneities, its lower growth rate renders amplification by SBS sensitive to parasitic instabilities, particularly forward Raman scattering and filamentation. These issues will be overestimated in simulations, but are still likely to be problematic in experiment and may ultimately limit achievable intensities and pulse compression. 

To summarize, though not all of the possible advantages of strongly-coupled stimulated Brillouin scattering are realizable and Raman amplifiers are likely to provide better performance in the regimes relevant for near-term applications, SC-SBS offers a promising alternative to SRS for plasma-based amplification and regimes exist where SBS is the only viable mechanism for plasma amplification. Conditions under which SC-SBS amplifiers may excel require further study.

\begin{acknowledgements}
This work was supported by NNSA Grant No.~DENA0002948, by AFOSR Grant No.~FA9550-15-1-0391, and by NSF Grant No.~PHY 1506372. M.R.E.~gratefully acknowledges the support of the NSF through a Graduate Research Fellowship. The presented simulations were performed at the High Performance Computing Center at Princeton University. The EPOCH code was developed as part of the UK EPSRC 300 360 funded project EP/G054940/1.
\end{acknowledgements}

\bibliography{References}

\end{document}